\documentclass[sigconf,nonacm]{acmart}

\usepackage{booktabs} 
\usepackage{amsmath,amssymb,amsfonts}
\usepackage{arydshln}
\usepackage{subfigure}
\usepackage{wrapfig,lipsum}
\usepackage{multirow}
\usepackage{caption}
\usepackage{hyperref}
\usepackage{algorithmic}
\usepackage{graphicx}
\usepackage{textcomp}
\usepackage{xcolor}

\setcopyright{acmcopyright}
\setcopyright{acmlicensed}
\setcopyright{rightsretained}

\acmPrice{15.00}

\begin{document}
\title{Comparative Analysis of Resource-Efficient CNN 
Architectures for Brain Tumor Classification}
\titlenote{This work has been accepted at IEEE ICCIT 2024. Please cite the version appearing in the ICCIT proceedings.}

\author{ \text{Md Ashik Khan}}
\affiliation{%
  \institution{Department of Computer Science and Engineering, \\Indian Institute of Technology Kharagpur}
  \state{West Bengal}
  \country{India}
}
\email{aasshhik98@gmail.com}

\author{ \text{Rafath Bin Zafar Auvee}}
\affiliation{%
  \institution{Department of Computer Science and Engineering,\\Bangladesh Army University of Science and Technology}
  \state{Saidpur}
  \country{Bangladesh}
}
\email{rafath.auvee@gmail.com}

\renewcommand{\shortauthors}{}

\begin{abstract}
Accurate brain tumor classification in MRI images is critical for timely diagnosis and treatment planning. While deep learning models like ResNet-18, VGG-16 have shown high accuracy, they often come with increased complexity and computational demands. This study presents a comparative analysis of effective yet simple Convolutional Neural Network (CNN) architecture  and  pre-trained ResNet18, and VGG16  model for brain tumor classification using two publicly available datasets: Br35H:: Brain Tumor Detection 2020 and Brain Tumor MRI Dataset. The custom CNN architecture, despite its lower complexity, demonstrates competitive performance with the pre-trained ResNet18 and VGG16 models. In binary classification tasks, the custom CNN achieved an accuracy of $\mathbf{98.67}$\% on the Br35H dataset and $\mathbf{99.62}$\% on the Brain Tumor MRI Dataset. For multi-class classification, the custom CNN, with a slight architectural modification, achieved an accuracy of $\mathbf{98.09}$\%, on the Brain Tumor MRI Dataset. Comparatively, ResNet18 and VGG16 maintained high performance levels, but the custom CNNs provided a more computationally efficient alternative. Additionally,the custom CNNs were evaluated using few-shot learning (0, 5, 10, 15, 20, 40, and 80 shots) to assess their robustness, achieving notable accuracy improvements with increased shots. This study highlights the potential of well-designed, less complex CNN architectures as effective and computationally efficient alternatives to deeper, pre-trained models for medical imaging tasks, including brain tumor classification. This study underscores the potential of custom CNNs in medical imaging tasks and encourages further exploration in this direction.
\end{abstract}

\keywords{Brain Tumor Classification, Convolutional Neural Network, Deep Learning, ResNet18, VGG16, Br35H , Brain Tumor MRI Dataset, Brain Tumor binary-classification CNN , Brain Tumor Multi-class-classification CNN , zero-shot Learning, few-shot Learning}

\maketitle

\section{Introduction}

Brain tumors are a significant health concern that requires accurate and fast diagnosis for optimal treatment. Timely and accurate diagnosis is crucial for effective treatment and improved outcomes. Magnetic Resonance Imaging (MRI) is a powerful diagnostic tool for identifying and classifying brain tumors with greater precision by providing detailed images of brain structures. However, the diagnostic process can be complex and time-consuming, requiring expertise in tumor diagnosis. 

In recent years, Deep learning algorithms have changed medical imaging by improving accuracy and efficiency in tumor detection and categorization. By applying these techniques to MRI data, researchers and clinicians can improve healthcare research and provide effective solutions for patients. Deep learning models such as ResNet18 \cite{resnet} and VGG16 \cite{vgg} have demonstrated remarkable performance in various image classification tasks. However, these models are often computationally intensive and require substantial resources, which may not be feasible in all clinical settings. 

The field of medical imaging has seen a surge in the application of deep learning techniques for various diagnostic tasks. For instance, Ronneberger et al. \cite{Ronneberger2015} introduced the U-Net architecture for biomedical image segmentation, which has been widely adopted for its efficiency and accuracy. Similarly, Litjens et al. \cite{litjens2017survey} provided a comprehensive review of deep learning in medical image analysis, highlighting the potential of these methods in improving diagnostic accuracy.

\begin{figure}[!ht]
\centering
\subfigure[Binary Brain Tumor Sample]{\includegraphics[width=0.2\textwidth]{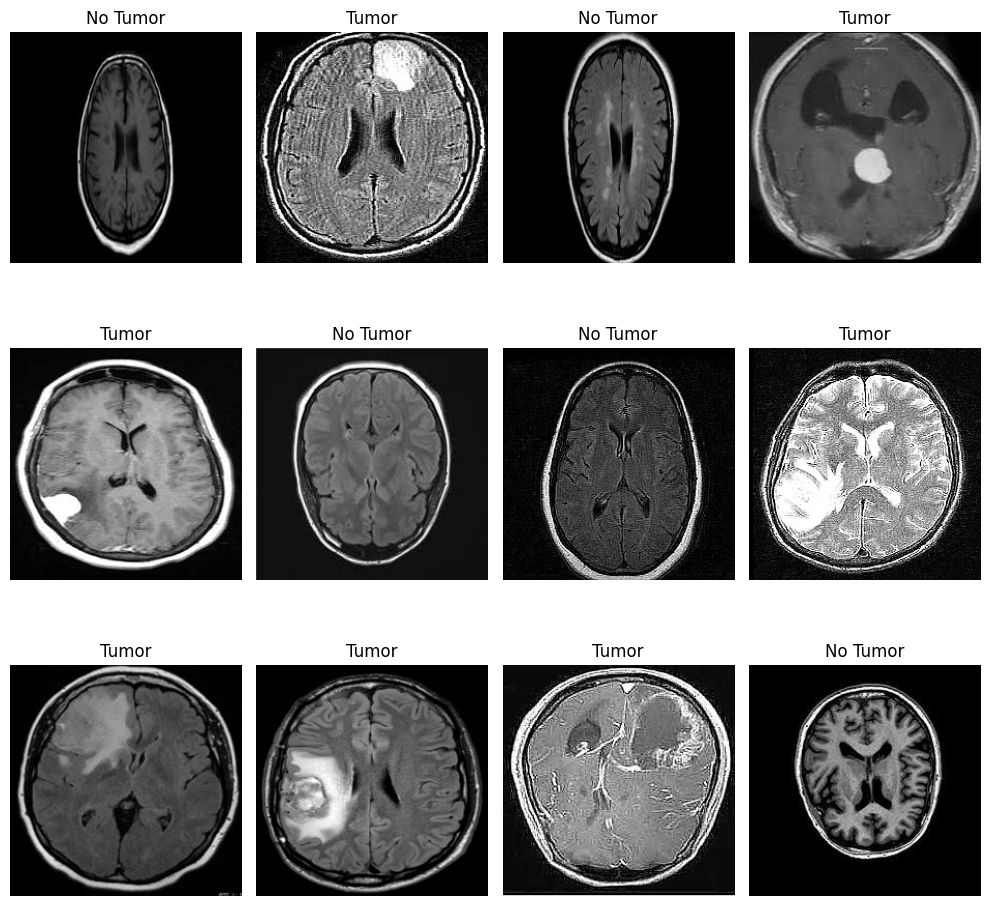}}
\subfigure[Multi-class Brain Tumor Sample]{\includegraphics[width=0.2\textwidth]{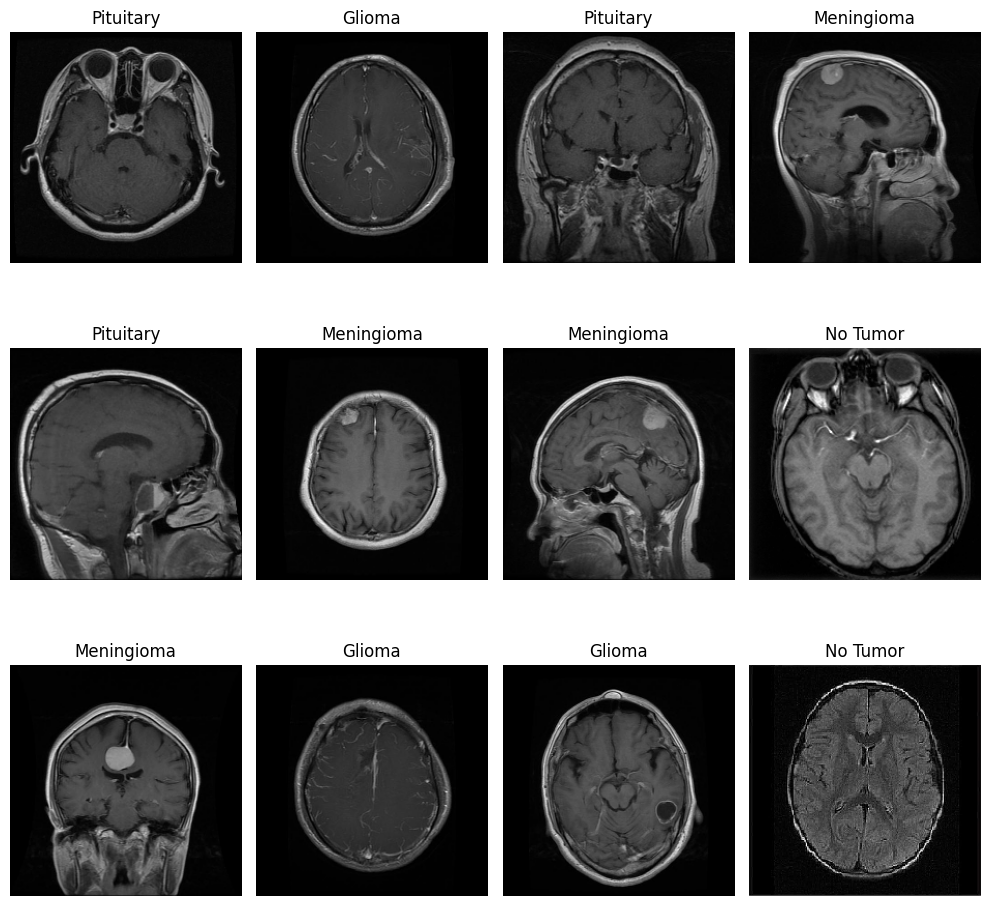}}
\caption{Brain Tumor MRI sample images.}
\label{fig:sample_images}
\end{figure}

In the context of brain tumor classification, several studies have explored the use of deep learning models. For example, Havaei et al. \cite{Havaei2017} proposed a novel CNN architecture for brain tumor segmentation, demonstrating high accuracy and robustness. Similarly, Pereira et al. \cite{Pereira2016} developed a deep learning model for brain tumor segmentation and classification, showing promising results in differentiating between various types of brain tumors.

However, the computational demands of these models can be a significant barrier to their widespread adoption, especially in resource-constrained environments. This has led to the development of more efficient architectures. For instance, Howard et al. \cite{Howard2017} introduced MobileNets, a family of lightweight CNN models designed for mobile and embedded vision applications. These models achieve high accuracy with significantly reduced computational complexity, making them suitable for deployment in resource-constrained settings.

This has led to an interest in developing simpler, custom Convolutional Neural Network (CNN) architectures that can deliver comparable performance with reduced computational complexity. Custom CNN architectures can potentially provide a balance between accuracy and computational efficiency, making them suitable for resource-constrained environments. These custom models are particularly advantageous in scenarios where computational resources are limited, as they require fewer parameters and simpler operations compared to their deeper counterparts \cite{Howard2017, Havaei2017}.

This study focuses on comparing the custom CNN architectures with pre-trained models such as ResNet18 \cite{resnet} and VGG16 \cite{vgg} For the work of classifying brain tumors using two in public datasets available: Br35H:: Brain Tumor Detection 2020 \cite{br35h2020}  and Brain Tumor MRI Dataset \cite{braintumormri2021}. The goal is to see if Simpler, less resource-intensive models can achieve competitive accuracy, thereby providing a viable alternative for resource-constrained environments. The study investigates the efficacy of custom CNNs in binary and multi-class classification tasks and compares their results against those obtained from ResNet18 \cite{resnet} and VGG16 \cite{vgg} as well as, their performance under few-shot learning conditions, where limited training data is available.

\begin{figure*}[!h]
    \centering
    \includegraphics[width=0.6\textwidth]{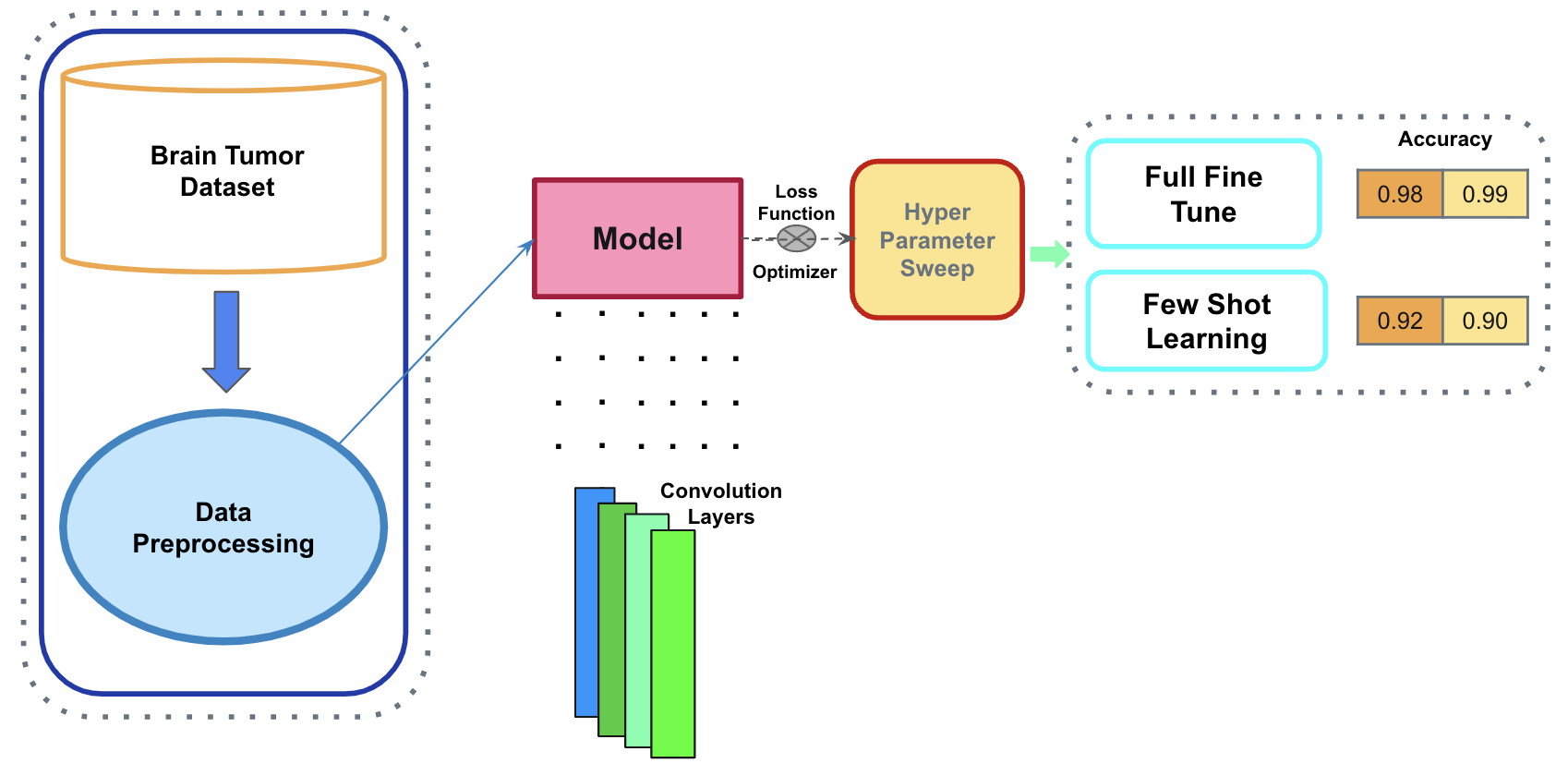}
    \caption{ \small Overview of the stages in the methodology: data preparation, model building, training and evaluation, and few-shot learning.}
    \label{fig:flowchart}
\end{figure*}

The findings from this study align with other research highlighting the effectiveness of simpler CNN models in various medical imaging tasks (e.g., \cite{kayalibay2017cnn}, \cite{reshi2021efficient}, \cite{zhang2017detecting}). This study's result demonstrates the potential of custom CNN architectures in medical imaging, offering a balance between accuracy and computational efficiency.

\section{Related Works}

Deep learning techniques have revolutionized medical image analysis, including brain tumor classification. Convolutional Neural Networks (CNNs) have been particularly successful in learning and extracting features from complex datasets, leading to improved diagnostic accuracy and efficiency.

Litjens et al. (2017) \cite{litjens2017survey} and Shen, Wu, and Suk (2017) \cite{shen2017deep} highlighted the significant advancements in deep learning for medical image analysis. CNNs have demonstrated remarkable performance in lesion detection, organ segmentation, and disease classification, including brain tumors. Pre-trained models, such as ResNet18 \cite{resnet} and VGG16 \cite{vgg}, have been widely adopted and achieved state-of-the-art results in various image recognition tasks.

In the context of brain tumor classification, custom CNN architectures have gained attention due to their potential for reduced computational complexity and comparable performance to pre-trained models. Recent studies have shown that customized CNNs, when combined with data augmentation and transfer learning techniques, can achieve competitive results with fewer layers and parameters\cite{alzoghby2023dual}.

Popular datasets for brain tumor classification include Br35H: Brain Tumor Detection 2020 \cite{br35h2020} and the Brain Tumor MRI Dataset \cite{braintumormri2021}. These datasets provide diverse MRI images with annotations, enabling precise training and evaluation of deep learning models \cite{menze2015multimodal}. Several studies have explored the potential of custom CNNs and pre-trained models in medical imaging. For example, Deepak and Ameer (2019) \cite{Deepak2019} utilized pre-trained models for brain tumor classification, achieving high accuracy. Similarly, Hemanth et al. (2017) \cite{Hemanth2017} demonstrated the effectiveness of CNNs in brain tumor segmentation.

In brain tumor classification, various techniques have been employed to enhance model performance. Kamnitsas et al. (2017) \cite{Kamnitsas2017} developed a multi-scale 3D CNN combined with a fully connected Conditional Random Field (CRF) for accurate brain lesion segmentation. Zhou et al. (2019) \cite{Zhou2019} provided a comprehensive review of deep learning techniques, including CNNs, for brain tumor segmentation, highlighting various approaches and their performance. Myronenko (2018) \cite{Myronenko2018} introduced a novel approach using an autoencoder for regularization in 3D MRI brain tumor segmentation, demonstrating improved performance. Isensee et al. (2018) \cite{Isensee2018} showed that existing CNN architectures can be effectively used for brain tumor segmentation without the need for new architectures. 

Transfer learning and data augmentation techniques have been widely used to enhance the performance of custom CNNs. Wang et al. (2019) \cite{GWang2019} utilized cascaded anisotropic CNNs for automatic brain tumor segmentation, achieving high accuracy. Hao et al. (2020) \cite{Hao2020} proposed a novel multi-scale CNN approach for automated brain tumor segmentation, further improving model performance.

At the same time, few-shot learning has emerged as a crucial method for training models with limited data, addressing a common constraint in medical imaging. Researchers have applied few-shot learning to brain tumor classification, demonstrating that custom CNNs can effectively learn and generalize even with minimal training data  \cite{achmamad2022few}. For instance, Antoniou et al. (2018) \cite{Antoniou2018} developed a few-shot learning approach that significantly improved model performance with limited data in various medical imaging tasks.


In addition to brain tumor classification, deep learning models have been applied to various other medical imaging tasks. Ronneberger et al. (2015) \cite{Ronneberger2015} introduced the U-net architecture, which has been widely used for biomedical image segmentation. SanaUllah et al. (2019) \cite{Khan2019} developed a novel deep learning-based framework for breast cancer classification, achieving high accuracy with reduced computational complexity.

The effectiveness of CNNs extends beyond brain tumor classification. Saira Charan et al. (2018) \cite{Charan2018} demonstrated the use of CNNs for breast cancer detection in mammogram images, achieving significant improvements in classification accuracy. Yu Gu et al. (2018) \cite{YGu2018} applied deep learning to lung nodule detection, further highlighting the versatility and potential of CNNs in various medical imaging tasks.

The integration of deep learning techniques in medical imaging has also been explored in the context of organ segmentation. Baid et al. (2021) \cite{Baid2021} presented the RSNA-ASNR-MICCAI BraTS 2021 benchmark for brain tumor segmentation and radiogenomic classification, showcasing the advancements in this field.

This study builds upon these related works, focusing on the comparative analysis of resource-efficient custom CNNs and their comparison with pre-trained models like ResNet18 \cite{resnet} and VGG16 \cite{vgg}. By exploring the potential of few-shot learning for efficient and effective brain tumor classification, this study aims to contribute to the development of practical, computationally efficient models for medical imaging tasks.
\\

\section{Methodology}

Following the example of other successful architectures discussed in the previous section, we use deep learning algorithms to classify brain tumors from two Brain Tumor MRI Datasets. We develop custom Convolutional Neural Network (CNN) architectures and evaluate the performance of them. To find their stand we compare them against pre-trained models. As shown in Figure \ref{fig:flowchart}, the methodology has four major stages: data preparation, model building,  training and evaluation, and few-shot learning experiments.

\subsection{Dataset}

\begin{itemize}
\item \textbf{ Br35H :: Brain Tumor Detection 2020}
\\The \href{https://www.kaggle.com/datasets/ahmedhamada0/brain-tumor-detection}{Br35H dataset } \cite{br35h2020}, created by Ahmed Hamada, focuses on binary classification problems, consisting of MRI pictures of brain scans with or without brain tumors. The dataset consists of $3000$ images, evenly distributed between the brain tumor and no brain tumor categories.

\begin{table}[!h]
  \centering
  \caption{\small The Overview of Brain Tumor MRI Datasets}
  \label{tab:dataset_table}
  \begin{tabular}{|c|c|c|}
    \hline
    \textbf{Dataset} & Br35H & Brain Tumor MRI \\
    \hline
    \textbf{Classes} & $2$ & $4$ \\
    \hdashline
    \textbf{Training Data} & $2400$ & $5712$ \\
    \hdashline
    \textbf{Testing Data} & $600$ & $1311$ \\
    \hdashline
    \textbf{Total Data} & $3000$ & $7023$ \\
    \hline
  \end{tabular}
\end{table}

\item \textbf{Brain Tumor MRI Dataset}
\\The \href{https://www.kaggle.com/datasets/masoudnickparvar/brain-tumor-mri-dataset}{Brain Tumor MRI Dataset} \cite{braintumormri2021}, created by Masoud Nickparvar, consists of MRI images categorized into four main classes based on the presence and type of brain tumor:
    \begin{itemize}
        \item \textbf{Glioma Tumor:} A type of tumor that originates in the glial cells of the brain, comprising 33\% of all primary brain tumors \cite{johnshopkins2021gliomas}.
        \item \textbf{Meningioma Tumor:} A typically slow-growing tumor that forms in the meninges, the protective layers surrounding the brain and spinal cord, accounting for about 30\% of all primary brain tumors \cite{johnshopkins2021meningioma}.
        \item \textbf{Pituitary Tumor:} A tumor that develops in the pituitary gland, located at the base of the brain, accounts for 9\% to 12\% of all primary brain tumors \cite{abta2021pituitary}.
        \item \textbf{No Tumor:} MRI images with no presence of a brain tumor.
    \end{itemize}
\end{itemize}

The dataset contains a total of $7023$ images, with each class having the following number of images: $1621$ glioma, $1645$ meningioma, $1757$  pituitary, and $2000$ no tumor. Notably, the no-tumor class images in this dataset were taken from the Br35H dataset. The images are labeled with clear annotations of their respective tumor types for training and evaluation purposes.

\begin{figure}[!h]
    \centering
    \includegraphics[width=0.45\textwidth]{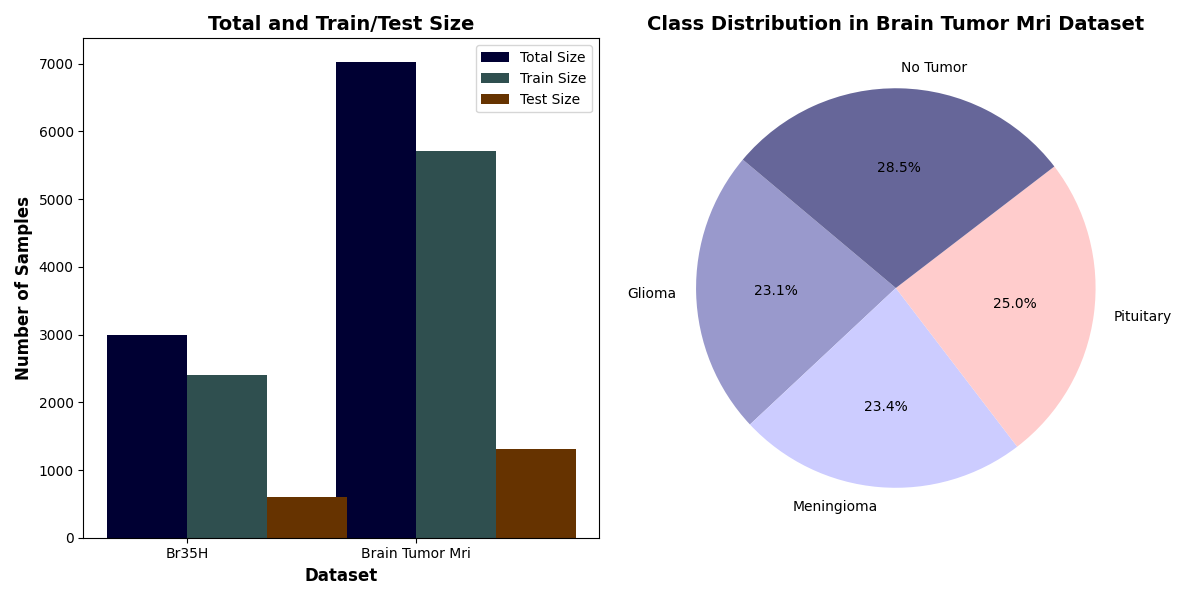}
    \caption{\small A summary of the training and testing data distribution for the Br35H and Brain Tumor MRI Dataset, along with the class distribution of MRI images within the Brain Tumor MRI Dataset.}
    \label{fig:data_distribution_table}
\end{figure}

\subsection{Data Pre-processing}
To ensure the data is suitable for training and evaluating our models, we applied several preprocessing steps:
\begin{itemize}
    \item All images were resized to $224\times224$ pixels to ensure uniformity and reduce computational complexity across the dataset. This size is commonly used for many pre-trained models like ResNet18 \cite{resnet}  and VGG16 \cite{vgg}.
    \item All images were converted to grayscale and blurred to eliminate noise and improve feature extraction during model training.
    \item The dataset was split into training and testing sets using an $80/20$ split, with a fixed random seed of $42$ to ensure reproducibility. Data loaders for both training and testing sets were created, with a batch size of $32$.To provide variability during the model training phase, shuffling was enabled for the training set.
\end{itemize}

\subsection{CNN Architecture}

For the task of brain tumor classification, two custom convolutional neural network (CNN) architectures were developed: 
\begin{itemize}
    \item \textbf{Two-layer CNN for binary classification}
    \\The Brain Tumor binary classification CNN (BTBCNN) consists of the following two layers:
    \begin{itemize}
        \item Input Layer: Accepts MRI images resized to $224\times224$ pixels.
        \item Convolutional Layers: Two convolutional layers with Batch Normalization and ReLU \cite{relu} activation functions, followed by max-pooling layers to reduce spatial dimensions.
            \begin{itemize}
                \item Conv1: $32$ filters, kernel size $3\times3$, padding $1$
                \item Conv2: $64$ filters, kernel size $3\times3$, padding $1$
            \end{itemize}
        \item Fully Connected Layers: One dense layer with ReLU \cite{relu} activations followed by a final output layer.
            \begin{itemize}
                \item FC1: $512$ units, followed by ReLU activation and dropout $(0.5)$
                \item FC2: $1$ unit (output layer)
            \end{itemize}
    \end{itemize}
    
    A dropout\cite{dropout} layer with a rate of $0.5$ is applied after the fully connected layer to prevent overfitting. This simple architecture was chosen for binary classification due to the smaller number of classes, allowing the model to learn and generalize effectively without requiring excessive complexity.
    \item \textbf{Three-layer CNN for multi-class classification}
    \\For the more complex multi-class classification task, a three-layer  Brain Tumor Multi-class-classification CNN (BTMCNN) was employed. The additional convolutional layer enhances the model's capacity to learn intricate features and patterns from the input images, which is essential for distinguishing between more than two classes. The architecture follows a similar design as the BTBCNN, with an extra convolutional layer:
    \begin{itemize}
       \item Additional Convolutional Layer with Batch Normalization and ReLU activation functions, followed by max-pooling layer.
        \begin{itemize}
            \item Conv3: $128$ filters, kernel size 3x3, padding 1
        \end{itemize}
    \end{itemize}
\end{itemize}

The choice of a dropout rate of $0.5$ in both architectures aims to strike a balance between reducing overfitting and maintaining the representation power of the neural networks. This value was selected based on its widespread use and effectiveness in preventing overfitting in various deep-learning applications.\\

In addition to the custom CNN architectures (BTBCNN and BTMCNN), two well-known pre-trained models, ResNet18 \cite{resnet} and VGG16 \cite{vgg}, were used for comparison purposes:

\begin{itemize}
    \item \textbf{ResNet18:} ResNet18 is a residual network with 18 layers, which employs residual connections to alleviate the vanishing gradient problem and facilitate the training of deeper networks. It contains 8 basic residual blocks, each consisting of 2 convolutional layers, 2 batch normalization layers, 2 ReLU activation layers, and 1 shortcut connection \cite{resnet}. 
    \item \textbf{VGG16:} VGG16, on the other hand, is a convolutional neural network with 16 weight layers, known for its simplicity and effectiveness in various image classification tasks. It comprises 13 convolutional layers, 5 max-pooling layers, and 3 fully connected layers, summing up to 43 layers in total \cite{vgg}.
\end{itemize}

 These models were chosen for their proven performance in image classification tasks, providing a valuable benchmark for evaluating the custom CNN architectures in the context of brain tumor classification.

\subsection{Training and Evaluation}

The Brain Tumor Binary Classification CNN (BTBCNN) and the Brain Tumor Multi-Class Classification CNN (BTMCNN) were trained using carefully selected loss functions and optimizers, along with a comprehensive hyperparameter sweep to ensure optimal performance. The training procedure involved data loading, model training, and evaluation using metrics such as accuracy, precision, recall, and F1-score.

For the binary classification task, BTBCNN was trained using Binary Cross-Entropy Loss with Logits (BCELosswithLogits) \cite{BCELoss}  as the loss function and the Adam \cite{adam} optimizer to update the model weights. For the multi-class classification task, BTMCNN was trained using Cross-Entropy Loss \cite{CrossEntropyLoss}  as the loss function and the Adam optimizer. The Cross-Entropy Loss effectively handles multiple classes, while the Adam optimizer ensures efficient and stable training.

A comprehensive hyperparameter sweep using a constant set of Learning rates [$0.001,0.0005,0.0001,0.00005$] was conducted to identify the optimal learning rate for both tasks. This range of learning rates covers different orders of magnitude, allowing for a balance between fast convergence and stable training. The best-performing learning rate was used for the full fine-tuning of the models.

This structured approach ensured that both models were effectively trained, yielding high performance in brain tumor classification tasks.

\subsection{Few Shot Learning}

To assess the robustness and adaptability of the models in scenarios with limited labeled data, a few-shot learning procedure was employed. This approach involved using the best-performing learning rate identified from the full fine-tuning process and applying it to smaller subsets of the training data. Steps:

\begin{itemize}
    \item The optimal learning rate from the full fine-tuning process was chosen for the few-shot learning experiments.
    \item To evaluate the initial performance and adaptability of the models, their accuracy was also assessed without any additional training (zero-shot learning).
    \item Subsets of the training data were created, ensuring that each subset maintained a balanced representation of every class:
    \begin{itemize}
	\item  5-shot: Containing $5$ samples per class.
	\item  10-shot: Containing $10$ samples per class.
        \item  15-shot: Containing $15$ samples per class.
        \item  20-shot: Containing $20$ samples per class.
        \item  40-shot: Containing $40$ samples per class.
        \item  80-shot: Containing $80$ samples per class.
    \end{itemize}
    Table \ref{tab:subset_few_shot} provides a comprehensive overview of the training data subsets used for few-shot learning experiments. 
    \item The models were trained on these reduced datasets using the selected learning rate.
\end{itemize}

\begin{table}[!ht]
\centering
\caption{\small Overview of training data subsets for Few-shot Learning }
\label{tab:subset_few_shot}
\label{tab}
\resizebox{0.49\textwidth}{!}{
    \begin{tabular}{c|cccccc}
    \hline
    \multirow{2}{*}{\textbf{Dataset}} & \multicolumn{6}{c}{\textbf{Samples per Subset}}  \\ \cline{2-7}
    & \textbf{5-shot} & \textbf{10-shot} & \textbf{15-shot} & \textbf{20-shot} & \textbf{40-shot} & \textbf{80-shot} \\
    \hline
    \textbf{Br35H} & $10$ & $20$ & $30$ & $40$ & $80$ & $160$ \\
    \hdashline
    \textbf{Brain Tumor MRI} & $10$ & $20$ & $30$ & $40$ & $80$ & $160$ \\
    \hdashline
    \multirow{1}{*}{\textbf{Brain Tumor MRI}} & \multirow{2}{*}{$20$} & \multirow{2}{*}{$40$} & \multirow{2}{*}{$60$} & \multirow{2}{*}{$80$} & \multirow{2}{*}{$160$} & \multirow{2}{*}{$320$}\\
    \textbf{(Multi-class)} & & & &&& \\
    \hline
    \end{tabular}
    }
\end{table}

The goal of this procedure was to test the models' ability to learn and generalize from limited data, simulating real-world scenarios where extensive labeled datasets may not be available. By evaluating the models under these conditions, we can better understand their potential for practical applications in brain tumor classification with limited resources.

\begin{figure*}[!ht]
\centering
\subfigure[BTBCNN for Br35H Dataset]{\includegraphics[width=0.28\textwidth]{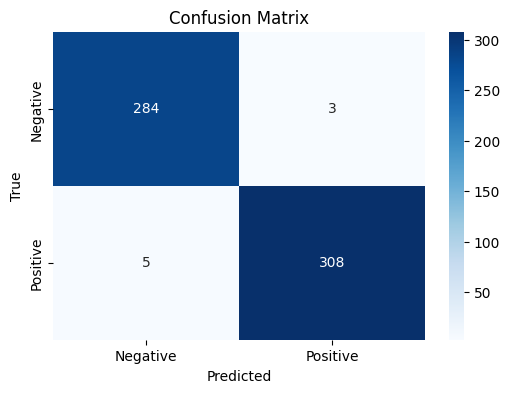}}
\subfigure[BTBCNN for Brain Tumor MRI Dataset ]{\includegraphics[width=0.28\textwidth]{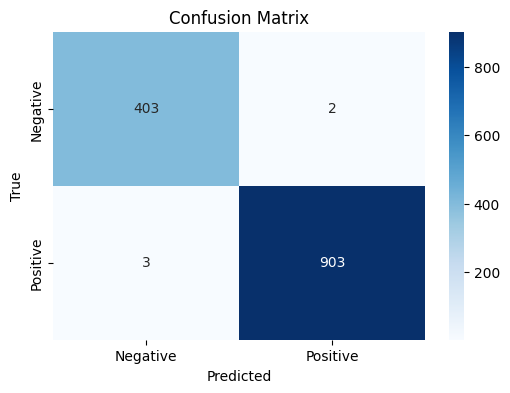}}
\subfigure[BTMCNN for Brain Tumor MRI Dataset]{\includegraphics[width=0.28\textwidth]{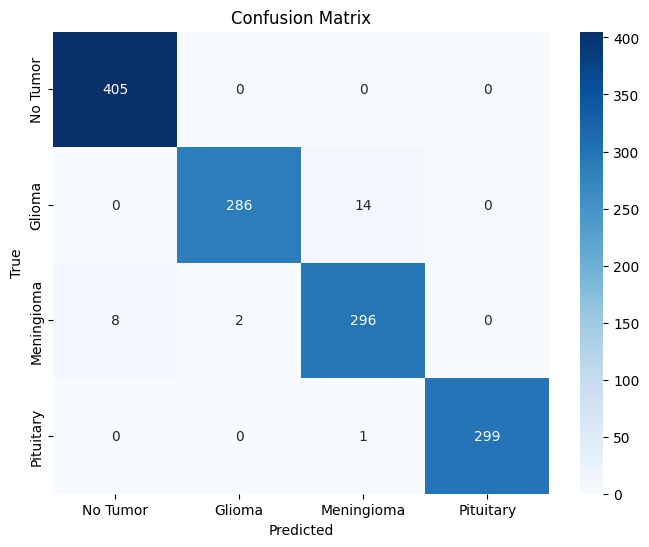}}
\caption{Confusion matrices  for binary classification using BTBCNN on the Br35H and Brain Tumor MRI Dataset, and for multi-class classification using BTMCNN on the Brain Tumor MRI Dataset of the Full-Finetuning Experiment.}
\label{fig:confusion_mat_fft}
\end{figure*}

\section{Results \& Analysis}

\subsection{Hyperparameter Sweep}
The hyperparameter sweep for the binary and multi-class brain tumor classification tasks involved evaluating different learning rates [$0.001, 0.0005, 0.0001, 0.00005$] to identify the optimal learning rate for each dataset. The following Table \ref{tab:hyper_param_br35H}, \ref{tab:hyper_param_MRI}, \ref{tab:hyper_param_MRI_Multiclass} and \ref{tab:hyper_param_accuracy} summarize the results of these experiments.

\begin{table}[!h]
\centering
\caption{\small Hyperparameter Sweep Results for Binary Classification Results for Br35H :: Brain Tumor Detection 2020 Dataset}
\label{tab:hyper_param_br35H}
\begin{tabular}{l|cccc}
\hline
\textbf{LR} & $0.001$ & $0.0005$ & $0.0001$ & $0.00005$ \\
\hline
\textbf{TP} & $307$ & $308$ & $302$ & $300$ \\
\textbf{TN} & $276$ & $284$ & $280$ & $280$ \\
\textbf{FP} & $11$ & $3$ & $7$ & $7$ \\
\textbf{FN} & $6$ & $5$ & $11$ & $13$ \\ \hdashline
\textbf{Precision} & $96.54\%$ & $99.03\%$ & $97.73\%$ & $97.74\%$ \\ \hdashline
\textbf{Recall} & $98.09\%$ & $98.40\%$ & $96.50\%$ & $95.84\%$ \\ \hdashline
\textbf{F1-Score} & $97.13\%$ & $98.71\%$ & $97.11\%$ & $96.78\%$ \\ \hdashline
\textbf{Accuracy} & $97.17\%$ & $\textbf{98.67}\%$ & $97.00\%$ & $96.67\%$ \\ 
\hline
\end{tabular}
\end{table}

The best-performing learning rate for the Br35H dataset was found to be $0.0005$, achieving the highest accuracy of $98.67\%$.

\begin{table}[!ht]
\centering
\caption{\small Hyperparameter Sweep Results for Binary Classification on Brain Tumor MRI Dataset}
\label{tab:hyper_param_MRI}
\label{tab}
    \begin{tabular}{l|cccc}
    \hline
    \textbf{LR} & $0.001$ & $0.0005$ & $0.0001$ & $0.00005$ \\
    \hline
    \textbf{TP} & $904$ & $901$ & $902$ & $901$ \\
    \textbf{TN} & $398$ & $404$ & $403$ & $405$ \\
    \textbf{FP} & $7$ & $1$ & $2$ & $0$ \\
    \textbf{FN} & $2$ & $5$ & $4$ & $5$ \\ \hdashline
    \textbf{Precision} & $99.23\%$ & $99.89\%$  & $99.78\%$  & $100\%$ \\ \hdashline
    \textbf{Recall} & $99.78\%$  & $99.45\%$  & $99.56\%$  & $99.45\%$  \\ \hdashline
    \textbf{F1-Score} & $99.50\%$  & $99.67\%$  & $99.67\%$  & $99.72\%$  \\ \hdashline
    \textbf{Accuracy} & $99.31\%$  & 99.54\%  & $99.56\%$  & $\textbf{99.62}\%$ \\ 
    \hline
    \end{tabular}
\end{table}

The optimal learning rate for the Brain Tumor MRI dataset was 0.00005, achieving an accuracy of 99.62\%.
For multi-class classification, the optimal learning rate was $0.0005$ with an accuracy of $98.09\%$ highlighted in \ref{tab:hyper_param_MRI_Multiclass}.

\begin{table}[!ht]
    \centering
    \caption{\small Hyperparameter Sweep Results for Multi-Class Classification on Brain Tumor MRI Dataset}
    \label{tab:hyper_param_MRI_Multiclass}
    \begin{tabular}{c|c|c|c|c}
    \hline
    \textbf{LR} & \textbf{Accuracy} & \textbf{Precision} & \textbf{Recall}  & \textbf{F1-Score} \\
    \hline
    $0.001$         & $97.10\%$   & $97.30\%$ & $97.20\%$ &  $97.25\%$ \\ \hdashline
    $0.0005$        & $\textbf{98.09}\%$    & $98.20\%$   & $98.10\%$ & $98.15\%$  \\ \hdashline
    $0.0001$        & $97.94\%$  & $98.00\%$   & $97.90\%$ & $97.95\%$  \\ \hdashline
    $0.00005$       & $97.56\%$  & $97.70\%$     & $97.60\%$ & $97.65\%$  \\ \hline
    \end{tabular}
\end{table}

The precision, recall, and F1-score values presented in table \ref{tab:hyper_param_br35H}, \ref{tab:hyper_param_MRI} and \ref{tab:hyper_param_MRI_Multiclass} demonstrate the effectiveness of the chosen learning rates in achieving high performance across the datasets. \\


\begin{table}[!ht]
  \centering
  \caption{\small Summary of Hyperparameter Sweep Results}
  \label{tab:hyper_param_accuracy}
  \resizebox{0.48\textwidth}{!}{
  
  \begin{tabular}{|c|c|c|c|c|}
    \hline
    \textbf{Dataset} & \textbf{Lr = $0.001$} & \textbf{Lr=$0.0001$} & \textbf{Lr=$0.0005$} & \textbf{Lr=$0.00005$} \\
    \hline
    \textbf{Br35H} & $97.17\%$ & $\textbf{98.67}\%$ & $97.00\%$ & $96.67\%$ \\
    \hdashline
    \textbf{Brain Tumorn MRI} & $99.31\%$ & $99.54\%$ & $99.54\%$ & $\textbf{99.62\%}$ \\
    \hdashline
    \multirow{1}{*}{\textbf{Brain Tumor MRI}} & \multirow{2}{*}{$97.10\%$} & \multirow{2}{*}{$\textbf{98.09\%}$} & \multirow{2}{*}{$97.94\%$} & \multirow{2}{*}{$97.56\%$} \\
    \textbf{(Multi-class)} & & & & \\
    \hline
  \end{tabular}
  }
\end{table}

In the Br35H dataset, BTBCNN with the learning rate of $0.0005$ achieved the best accuracy of $98.67$\% with the highest precision ($99.03\%$), recall ($98.40\%$), and F1-score ($98.71\%$), ensuring minimal false positives and false negatives. Similarly, in the Brain Tumor MRI dataset, the learning rate of 0.00005 showcased exceptional performance with perfect precision ($100\%$) and a high F1-score ($99.72\%$) indicating the BTBCNN's accurate classification of brain tumors without any false positives. Lastly, for the multi-class Brain Tumor MRI (MAP) dataset, BTMCNN with the learning rate of $0.0005$ provided strong performance with high precision ($98.20\%$), recall ($98.10\%$), and F1-score ($98.15\%$), confirming effectiveness in handling multi-class brain tumor classification tasks. This can be observed by the figure \ref{fig:confusion_mat_fft}, which provides a visual representation of the model's classification accuracy.

\subsection{ Full Fine-Tuning of Custom CNNs and Pretrained Models}

Upon examination of Table \ref{tab:cnn_resnet_vgg}, it becomes apparent that both custom models (BTBCNN \& BTMCNN) and pre-trained models (ResNet18 \& VGG16) demonstrated excellent performance in brain tumor classification based on full fine-tuning experiments. For binary classification on the Br35H dataset, BTBCNN achieved an accuracy of 98.67\%, while ResNet18 and VGG16 reached 99.33\% and 99.50\%, respectively. In terms of precision, recall, and F1-score, BTBCNN obtained 99.03\%, 98.40\%, and 98.71\%, whereas ResNet18 and VGG16 had slightly better results. On the Brain Tumor MRI dataset, BTBCNN reached an accuracy of 99.62\%, and both ResNet18 and VGG16 achieved 100\%. In multi-class classification on the Brain Tumor MRI dataset, BTMCNN had an accuracy of 98.09\%, ResNet18 of 99.69\%, and VGG16 of 99.39\%.\\



\begin{figure}[!h]
    \centering
    \includegraphics[width=0.48\textwidth]{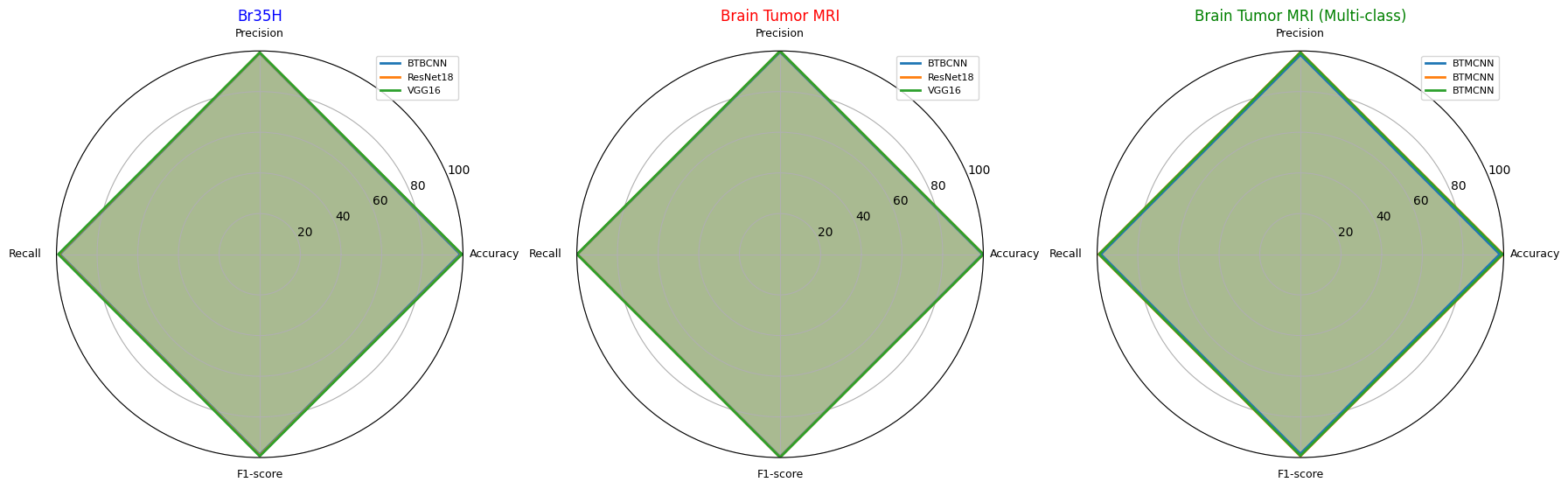}
    \caption{ Performance comparison of custom CNNs (BTBCNN, BTMCNN) and pre-trained models (ResNet18, VGG16) in terms of accuracy, precision, recall, and F1-scores for binary and multi-class classification tasks.}
    \label{fig:cnn_vs_pretrained_analysis}
\end{figure}


Figure \ref{fig:cnn_vs_pretrained_analysis} provides a comprehensive visualization of the consistently high performance achieved by all models across the datasets. When analyzing these results, it is evident that the slight variations in metrics are often within a fraction of a percent, making it challenging to distinguish a clear superior model based solely on these evaluation metrics.

\begin{table}[!h]
  \centering
  \caption{\small  Full Fine-Tuning Accuracies of Custom CNNs (BTBCNN \& BTMCNN), Pretrained ResNet18 and Pretrained VGG16}
  \label{tab:cnn_resnet_vgg}
  \begin{tabular}{|c|c|c|c|c|}
    \hline
    \textbf{Dataset} & \textbf{Custom CNN} & \textbf{ResNet 18} & \textbf{VGG16} \\
    \hline
    \textbf{Br35H} & $98.67$ & $99.33$ & $99.50$ \\
    \hdashline
    \textbf{Brain Tumor MRI} & $99.62$ & $100.00$ & $100.00$ \\
    \hdashline
    \multirow{1}{*}{\textbf{Brain Tumor MRI}} & \multirow{2}{*}{$98.09$} & \multirow{2}{*}{$99.69$} & \multirow{2}{*}{$99.39$} \\
    \textbf{(Multi-class)} & & & \\
    \hline
  \end{tabular}
\end{table}

The primary advantage of Custom CNNs lies in their simpler architecture and fewer parameters. For instance, BTBCNN has fewer layers and uses only two convolutional layers, while ResNet18 has $18$ layers (16 convolutional layers). BTMCNN has only $51,476,484$ parameters compared to VGG16's $138,357,544$. This reduced complexity is further emphasized by the training time comparison for $50$ epochs on the Brain Tumor MRI Dataset: BTMCNN, ResNet18, and VGG16 take respectively $550.37$, $569.48$ and  $2474.92$ seconds.\\

To further illustrate the efficiency of custom CNNs, we present the average inference time per batch (for batch size $128$) for different models on two datasets: Br35H and Brain Tumor MRI. The inference times are shown in the table \ref{tab:inference_time}.

\begin{table}[!ht]
\centering
\caption{Average Inference Time per Batch (Milliseconds) for Different Models}
\label{tab:inference_time}
\begin{tabular}{llccc}
\toprule
\multirow{2}{*}{\textbf{Dataset}} & \multirow{2}{*}{\textbf{Model} } & \multirow{1}{*}{\textbf{Average Inference Time }}  \\
 & &\textbf{per Batch (Milliseconds)}\\
\midrule
\multirow{4}{*}{Br35H} & BTBCNN & \textbf{0.9} \\
 & BTMCNN & 1.2\\
 & ResNet18 & 3.8 \\
 & VGG16 & 2.8 \\
\midrule
\multirow{4}{*}{Brain Tumor MRI} & BTBCNN & \textbf{0.9} \\
 & BTMCNN & 1.4 \\
 & ResNet18 & 4.0 \\
 & VGG16 & 2.8 \\
\bottomrule
\end{tabular}
\end{table}

These results demonstrate that custom CNNs, such as BTBCNN and BTMCNN, have significantly lower inference times compared to ResNet18 and VGG16. This efficiency makes custom CNNs a viable option for tasks where computational resources are limited or where faster training and inference times are desired.\\

In conclusion, while pre-trained models like ResNet18 and VGG16 show slightly higher accuracies and metrics, custom CNNs like BTBCNN and BTMCNN offer significant advantages in terms of simplicity, efficiency, and resource usage. These advantages make custom CNNs particularly suitable for environments with limited computational resources or where faster training and inference times are critical. By balancing accuracy with computational efficiency, custom CNNs provide a practical alternative to deeper, pre-trained models in medical imaging tasks, particularly in brain tumor classification.

\subsection{Few-shot Learning}


Few-shot learning experiments were conducted to assess the robustness and adaptability of the custom CNNs (BTBCNN \& BTMCNN). From Table \ref{tab:few_shot_accuracy_table} it is observed that 0-shot results show bias, while 5-shot to 80-shot demonstrate the custom CNN's strength and adaptability. The bias observed in 0-shot results can be attributed to the model's lack of exposure to the target dataset, causing it to rely on its initial random weights. The results indicate significant improvements in accuracy as the number of training samples increases, showcasing the effectiveness of our few-shot learning approach. For the BTBCNN model on the Brain Tumor Detection Dataset, the accuracy increased from 52.00\% to 80.17\% with an 80-shot dataset, representing a total improvement of 28.17\%. Notable jumps occur between 0-shot to 10-shot (11.50\%) and 10-shot to 15-shot (4.33\%). Similarly, the BTBCNN model on the Brain Tumor MRI Dataset showed an accuracy increase from 69.11\% to 89.78\% with an 80-shot dataset, representing a total improvement of 20.67\%. The most significant improvement is seen from 0-shot to 10-shot (13.80\%), with steady but smaller gains beyond 20-shot. The BTMCNN model on the Brain Tumor MRI Dataset demonstrated the highest improvement, increasing accuracy from 22.88\% to 78.03\% with an 80-shot dataset, representing a total improvement of 55.15\%. The largest improvement is between 5-shot and 10-shot (31.66\%), showing substantial learning capability with limited data. These findings underscore the robustness and adaptability of our custom CNNs, making them suitable for brain tumor classification tasks in real-world scenarios where labeled data might be limited. The models' ability to achieve high accuracy even with minimal training samples highlights the efficiency of our few-shot learning approach, particularly in the early stages.\\

\begin{table}[h!]
\centering
\caption{Overview of Few-shot Learning Results for BTBCNN and BTMCNN}
\label{tab:few_shot_accuracy_table}
\resizebox{0.49\textwidth}{!}{
  
    \begin{tabular}{ccccccccc}
    \toprule
    \textbf{Model} & \textbf{Dataset} & \textbf{0-shot} & \textbf{5-shot} & \textbf{10-shot} & \textbf{15-shot} & \textbf{20-shot} & \textbf{40-shot} & \textbf{80-shot} \\
    \midrule
    \multirow{2}{*}{BTBCNN} & Br35H & 52.00\% & 52.17\% & 63.67\% & 68.00\% & 70.83\% & 76.83\% & 80.17\% \\
     & Brain Tumor MRI & 69.11\% & 69.11\% & 82.91\% & 82.07\% & 85.05\% & 89.32\% & 89.78\% \\
    \midrule
    BTMCNN & Brain Tumor MRI & 22.88\% & 31.50\% & 63.16\% & 68.57\% & 70.40\% & 72.77\% & 78.03\% \\
    \bottomrule
    \end{tabular}
    }

\end{table}


  

\begin{figure}[!h]
    \centering
    \includegraphics[width=0.48\textwidth]{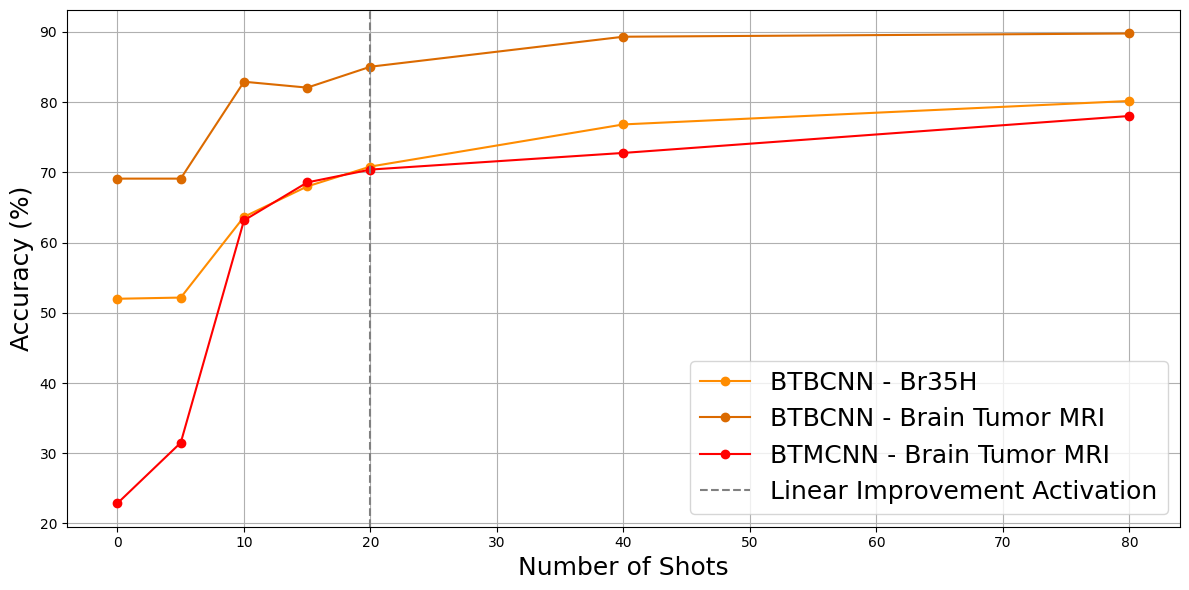}
    \caption{ Comparison of Different Few-shot Learning Accuracies of Custom CNNs}
    \label{fig:few_shot_accuracy_comp}
\end{figure}

Figure \ref{fig:few_shot_accuracy_comp} shows the progressive improvement in BTBCNN and BTMCNN accuracy as training samples increase. Both models demonstrate significant accuracy jumps from 0-shot to 10-shot. After 20 shots, the improvements become more linear, suggesting a diminishing return on accuracy gains as more data is introduced. This states the models' exceptional ability to adapt and learn effectively from limited data even without pretraining.




\section{Conclusion}

This study successfully developed and evaluated custom convolutional neural networks (BTBCNN and BTMCNN) to classify brain tumors.  Our findings show that these customized models maintaining a simpler design, achieve competitive accuracy compared to well-established architectures such as ResNet18 \cite{resnet} and VGG16 \cite{vgg} while offering substantial benefits in terms of computational efficiency. The Custom CNNs provide competitive accuracy in brain tumor classification. Their simpler design offers faster training and quicker inference times. They can be effectively adapted for tasks with limited data availability. The slight performance difference is offset by the advantages of faster training and inference times. However, this study has several limitations. The research relied on two publicly available datasets, which may not fully represent the diverse characteristics of brain tumors across different patient populations and imaging technologies. The models’ simplicity, while advantageous for computational efficiency, might restrict their ability to detect complex and nuanced features essential for intricate tumor classifications or diverse imaging scenarios.\\

Overall, the custom CNNs (BTBCNN and BTMCNN) present a balanced approach, achieving high accuracy with minimal complexity. This makes them a viable and efficient solution for brain tumor classification, particularly in resource-constrained settings. They offer a promising alternative for practical applications in medical image analysis. 

\vspace{30pt}

\bibliographystyle{ACM-Reference-Format}
\bibliography{ICVGIP-Latex-Template}

\appendix




\end{document}